\documentclass[manuscript,screen,nonacm]{acmart}

\AtBeginDocument{%
  }

\begin{document}

\title{When Interpretations Persist: Downstream Influence in Gaze-Aware Assistants}

\author{Mingcong Ding}
\authornote{Corresponding author.}
\email{mingcong.ding@tum.de}
\affiliation{%
  \institution{Human-Centered Technologies for Learning, Technical University of Munich}
  \city{Munich}
  \country{Germany}
}
\affiliation{%
  \institution{Munich Center for Machine Learning (MCML)}
  \city{Munich}
  \country{Germany}
}

\author{Chenyan Zhang}
\email{zhangchenyan2024@163.com}
\affiliation{%
  \institution{Technical University of Munich}
  \city{Munich}
  \country{Germany}
}

\author{Zhuojiang Cai}
\email{cai.zhuojiang@tum.de}
\affiliation{%
  \institution{Human-Centered Technologies for Learning, Technical University of Munich}
  \city{Munich}
  \country{Germany}
}
\affiliation{%
  \institution{Munich Center for Machine Learning (MCML)}
  \city{Munich}
  \country{Germany}
}

\author{Dongyang Guo}
\email{dongyang.guo@tum.de}
\affiliation{%
  \institution{Human-Centered Technologies for Learning, Technical University of Munich}
  \city{Munich}
  \country{Germany}
}
\affiliation{%
  \institution{Munich Center for Machine Learning (MCML)}
  \city{Munich}
  \country{Germany}
}

\author{Babette B\"uhler}
\email{babette.buehler@tum.de}
\affiliation{%
  \institution{Human-Centered Technologies for Learning, Technical University of Munich}
  \city{Munich}
  \country{Germany}
}
\affiliation{%
  \institution{Munich Center for Machine Learning (MCML)}
  \city{Munich}
  \country{Germany}
}

\author{Enkelejda Kasneci}
\email{enkelejda.kasneci@tum.de}
\affiliation{%
  \institution{Human-Centered Technologies for Learning, Technical University of Munich}
  \city{Munich}
  \country{Germany}
}
\affiliation{%
  \institution{Munich Center for Machine Learning (MCML)}
  \city{Munich}
  \country{Germany}
}

\renewcommand{\shortauthors}{Ding et al.}

%


\begin{abstract}
AI assistants may carry forward not only observations but also interpretations of what a person is doing or may do next across time. Once retained, these interpretations become context for later predictions and system-triggered outputs. We study whether this representation step changes later behavior in gaze-aware streaming assistance. Holding retrieved events, their order, and the target input fixed, we find that Free-form Interpretation amplifies historical influence in both directions: correct histories produce more helpful changes and wrong histories more harmful changes. Unsupported and broader interpretations about activity or future behavior are associated with greater harmful influence. Evidence-Bounded Interpretation preserves distinctions between observation, inference, uncertainty, and temporal scope, reducing unsupported and overcertain content and lowering wrong-source harmful influence while retaining most helpful influence. The same effect structure appears with a second model. These findings identify persistent interpretation as a consequential design choice for history-aware human-AI systems.

\end{abstract}



\begin{teaserfigure}
  \includegraphics[width=\textwidth]{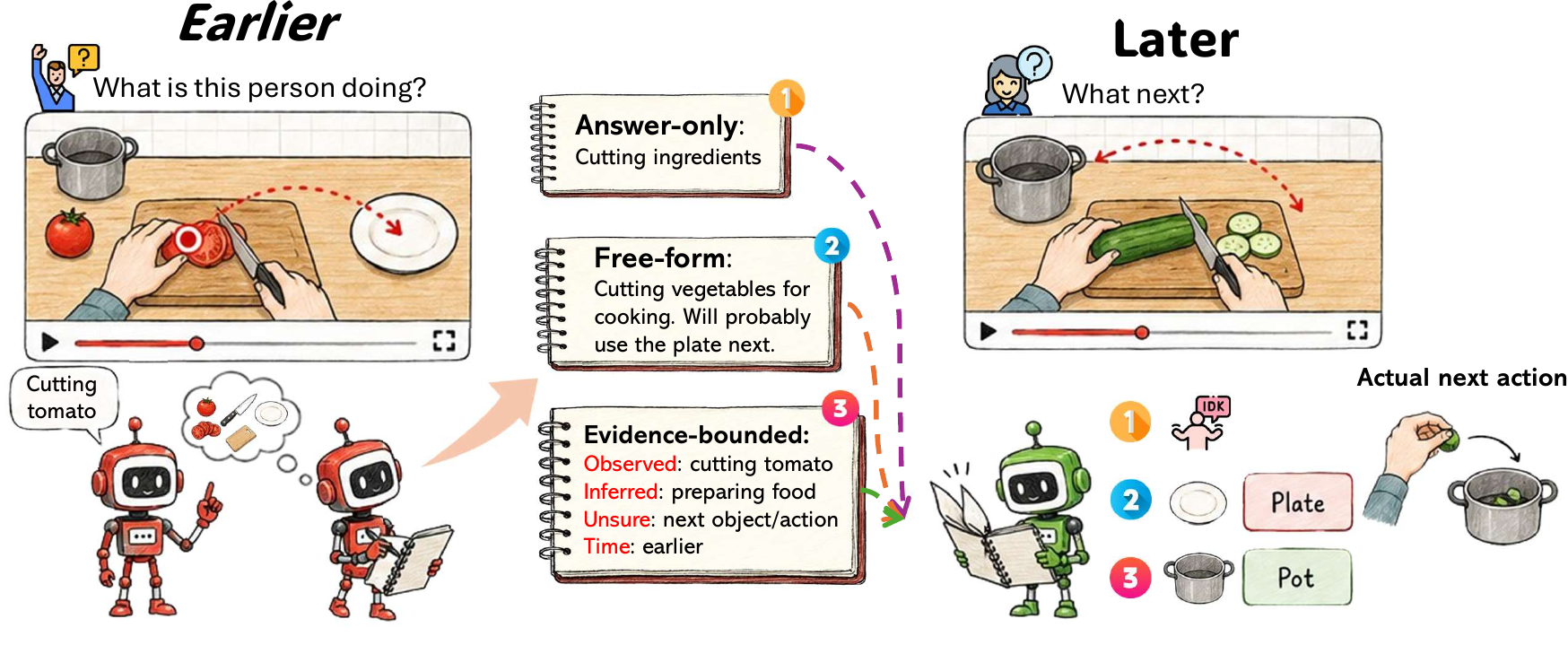}
  \caption{Illustrative example of persistent interpretation changing downstream influence. The same source-time evidence can be retained as a direct answer, rewritten as a fluent interpretation, or represented with explicit epistemic status before later reuse. The experiments use the original STREAMGAZE task formats.}
  \Description{A conceptual teaser showing that fluent persistent interpretations can amplify both helpful and harmful downstream influence, while Evidence-Bounded Interpretation preserves distinctions between observed, inferred, uncertain, and temporally scoped content.}
  \label{fig:teaser}
\end{teaserfigure}


\maketitle

\section{Introduction}

AI assistants that operate across an unfolding activity may need to use earlier information when deciding what to do next. A gaze-aware assistant, for example, may use earlier observations to recall scene history, predict a future action, or decide when to issue an alert. In such systems, what carries forward may be more than what was observed. The assistant may also retain its own interpretations of what a person was doing or might do next. Observing that a person looked at a tomato is different from retaining the interpretation that they were preparing to cut it. The latter turns a momentary inference about the person into context for later assistance.

We call a model-generated description or inference derived from behavioral evidence a \textbf{machine interpretation}, and a \textbf{persistent interpretation} when it is retained and later reused as context. Once persistent, an interpretation acquires a second function: it is no longer only an output about an earlier event, but also part of the information that shapes later system behavior. This makes persistent interpretation a design choice for history-aware AI systems. When the underlying behavioral evidence stays the same, does the way that evidence is turned into persistent interpretation change what the system later does?

Prior work shows that model-generated history and retrieved experience can improve later performance, but can also propagate errors when retained information is incorrect or misaligned~\cite{wu2024streambench,xiong2026memory}. Other work has identified failures in memory construction and updating, as well as representations that obscure the source or role of stored information~\cite{chen2025halumem,jin2026mitigating,xu2026memory,liao2026belief}. These findings establish that history matters, but leave a different step unresolved: whether transforming the same behavioral evidence into a fluent machine interpretation changes how strongly that history shapes subsequent decisions.

Gaze-aware assistance makes this representational question concrete. Video and gaze provide relatively direct evidence about visible objects and visual attention, while interpretations of ongoing activity or possible future behavior require additional inference~\cite{hayhoe2005eye,li2021eye,pani2026gaze,peng2026eye,lee2025streamgaze,lai2024listen}. The setting therefore lets us study what happens when statements that range from relatively observable evidence to broader interpretations about a person become persistent context for later assistance.

Using STREAMGAZE~\cite{lee2025streamgaze}, we hold the retrieved earlier events, their order, and the later target input fixed while varying the persistent representation constructed from those events. We find that \emph{Free-form Interpretation} acts as an influence amplifier. Relative to \emph{Answer Only}, correct source histories produce more helpful downstream changes, while wrong source histories produce more harmful changes. The same effect structure appears with GLM-4.6V-Flash~\cite{vteam2025glm45vglm41vthinkingversatilemultimodal}. Human analyses further show that harmful influence is associated with unsupported content and is higher for broader interpretations about activity or possible future behavior.

The two-sided effect makes the form of persistent interpretation a design question: richer interpretation can make useful history more useful while also making wrong history more consequential. We therefore ask what form an interpretation should take when it persists. We evaluate \emph{Evidence-Bounded Interpretation} (EB), a representation that keeps observation, inference, uncertainty, and temporal scope distinguishable. EB reduces unsupported and overcertain content and substantially lowers wrong-source harmful influence while retaining most of the helpful influence of Free-form Interpretation. Counterfactual correction further shows that the semantic content carried forward contributes to later failures. These results identify epistemic representation as part of the design of persistent machine interpretation.

Our contributions are:
\begin{itemize}
    \item We identify persistent interpretation as a consequential representation step in the reuse of behavioral information. Transforming the same earlier evidence into Free-form Interpretation increases downstream influence in both directions, producing more helpful changes from correct histories and more harmful changes from wrong histories. This amplification pattern is reproduced with a second model.
    \item We characterize what makes persistent interpretation more consequential. Even within wrong-source histories, unsupported interpretations are associated with substantially more harmful downstream changes, while a separate human audit shows higher harmful-flip rates as interpretations extend from relatively observable information toward activity and possible future behavior.
    \item We derive and test a concrete design response for persistent machine interpretation. Evidence-Bounded Interpretation keeps observation, inference, uncertainty, and temporal scope distinguishable, reduces unsupported and overcertain content, and substantially lowers wrong-source harmful influence while retaining most correct-source helpful influence. Counterfactual correction further shows that the semantic content carried forward contributes to later failures.
\end{itemize}

\section{Related Work}

\subsection{Gaze-Aware and Streaming Video Understanding}

Eye gaze provides information about visual attention during ongoing behavior~\cite{hayhoe2005eye,li2021eye}. Recent first-person and multimodal work uses gaze for collaborative video question answering~\cite{ilaslan2023gazevqa}, gaze anticipation~\cite{lai2024listen}, mistake detection~\cite{mazzamuto2025gazing}, and gaze-conditioned visual-language reasoning~\cite{pani2026gaze,peng2026eye}. STREAMGAZE combines gaze with a causal streaming protocol spanning object, temporal, future-action, and proactive tasks~\cite{lee2025streamgaze}. Related benchmarks such as StreamingBench, OVO-Bench, and OVBench study online video understanding under temporally arriving evidence~\cite{lin2026streamingbench,niu2025ovo,huang2025online}. Together, these works show how gaze-conditioned reasoning is expanding from first-person scene understanding toward temporally evolving and online video settings.

Our question concerns what happens after such evidence has already produced a model interpretation. In our study, we distinguish statements closely tied to visible objects or attended regions from statements that add activity or prospective inference, drawing on gaze and egocentric work that spans attention, action, and anticipation~\cite{hayhoe2005eye,li2021eye,lai2024listen,peng2026eye}. This gives us a controlled setting for studying how increasingly interpretive statements behave after persistence. The works above primarily evaluate judgments produced from the current stream; we study what happens when a model-generated interpretation is retained and later reused as context for another decision.

\subsection{Persistent Context, Memory Construction, and Experience Reuse}

Language agents increasingly externalize prior information into persistent state that can influence later reasoning. Generative Agents synthesizes higher-level reflections over accumulated observations~\cite{park2023generative}, while Reflexion and ExpeL reuse textual experience from previous attempts~\cite{shinn2023reflexion,zhao2024expel}. Long-term memory systems extend this idea through explicit storage and retrieval, including MemGPT, Agent Workflow Memory, A-MEM, ReasoningBank, and multimodal long-term memory~\cite{packer2023memgpt,wang2024agent,xu2026mem,ouyang2026reasoningbank,long2026seeing}. These systems establish that model-generated information can persist beyond the interaction in which it was produced.

Recent work also treats memory construction, retrieval, and management as distinct design problems. LongMemEval evaluates long-term interactive memory~\cite{wu2024longmemeval}, SeCom studies memory construction and retrieval for conversational agents~\cite{pan2025secom}, and MemInsight, REMem, and Memory-R1 examine memory augmentation, episodic reasoning, and memory management~\cite{salama2025meminsight,shu2026remem,yan2026memory}. A recent survey similarly describes a shift from simple storage toward systems that construct and use experience as reusable state~\cite{luo2026storage}. This literature shows that memory construction and selection can affect later use.

Persistent history can also introduce failure. StreamBench shows that self-generated experience can help or hurt subsequent performance depending on its correctness~\cite{wu2024streambench}, while memory management affects experience-following behavior~\cite{xiong2026memory} and memory systems can contain hallucinated content~\cite{chen2025halumem}. Building on this literature, we isolate a different step before reuse: whether the same earlier behavioral evidence has different downstream consequences after it is transformed from a direct answer into a fluent machine interpretation. Holding retrieved source events, their order, and the later target input fixed lets us study this representation transformation separately from retrieval quality or history availability.

\subsection{Uncertainty, Provenance, and Memory Reliability}

How uncertainty is represented can affect how AI outputs are interpreted and relied upon. Work on communicating uncertainty examines how uncertainty expressions shape judgments~\cite{van2019communicating}, while studies of human-AI decision making investigate the relationship between uncertainty presentation, trust, reliance, and calibration~\cite{salimzadeh2024dealing,schafer2025ai,kim2024m,ma2024you}. Linguistic calibration similarly addresses overconfident natural-language output~\cite{mielke2022reducing}. These studies concern human interpretation of AI uncertainty, whereas our experiments measure downstream model behavior rather than user trust or reliance. We use this literature as design context for preserving information status, not as evidence of user-level effects in our setting.

Related reliability problems arise when uncertain or model-generated information becomes persistent memory. Typed memory representations address failures to preserve the provenance and role of stored information~\cite{jin2026mitigating}, and memory provenance laundering examines amplification after the original status of persistent information becomes obscured~\cite{xu2026memory}. HaluMem studies hallucinated content in agent memory~\cite{chen2025halumem}, while Agentic Uncertainty Quantification considers uncertainty within agentic reasoning~\cite{zhang2026agentic}. Belief Memory is closely related to our design motivation: it treats memory under partial observability and preserves uncertainty rather than collapsing ambiguous evidence into one deterministic belief~\cite{liao2026belief}. Its representation uses multiple probabilistic candidate beliefs, whereas EB preserves evidence status within a structured textual interpretation.

These works establish uncertainty preservation, provenance, and structured memory as important concerns in persistent AI context. Our empirical contribution instead precedes the design response: we isolate the representation step and show that converting the same earlier behavioral evidence into fluent persistent interpretation changes its downstream influence. We then test whether preserving epistemic distinctions changes that influence when the interpretation is reused.

\section{Study Setting}

\begin{figure*}[t]
    \centering
    \includegraphics[width=\textwidth]{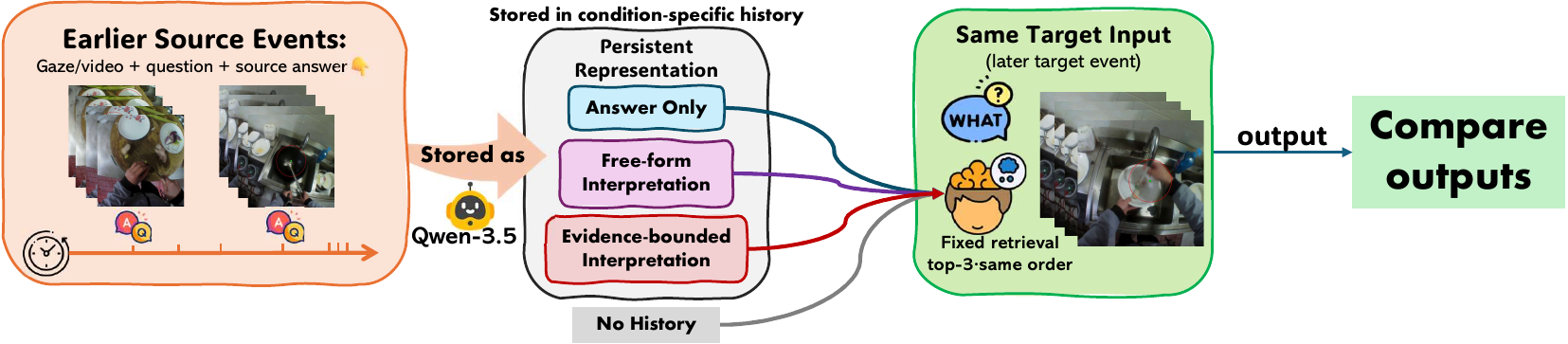}
    \caption{Controlled study setting. For each later target event, earlier source events are retrieved once and then reused across representation conditions. The retrieved events, their order, and the target input are held fixed; only the persistent representation changes. Wrong-source (WS) and correct-source (CS) analyses isolate histories in which all three retrieved direct source predictions are incorrect or correct, respectively.}
    \Description{Study setting showing three earlier source events from the same video followed by a later target event. The same retrieved source events branch into Answer Only, Free-form Interpretation, and Evidence-Bounded Interpretation conditions before being used with the same target input. Wrong-source and correct-source populations are evaluated using harmful and helpful downstream flips, respectively.}
    \label{fig:study-setting}
\end{figure*}

We use gaze-aware streaming video as a controlled case in which an AI system can reuse earlier interpretations about a person when making later decisions. We define a \emph{machine interpretation} as a model-generated description or inference derived from source-time behavioral evidence, and a \emph{persistent interpretation} as such an interpretation when it is retained and later reused as context. As illustrated in Figure~\ref{fig:study-setting}, we keep the retrieved source events, their order, and the target input fixed across conditions, while varying their persistent representation. Free-form Interpretation may add higher-level inference beyond the direct source answer, so the comparison evaluates the persistent representation as a whole.

\subsection{Persistence and Retrieval}

We use four STREAMGAZE tasks~\cite{lee2025streamgaze} as sources of earlier events: Object Identification Easy (OI-E), Object Identification Hard (OI-H), Object Attribute Recognition (OAR), and Non-Fixated Object Identification (NFI). The resulting source library contains 4,513 entries anchored to direct source predictions. We evaluate their downstream influence on six later target tasks: Object Transition Prediction (OTP), Scene Recall (SR), Gaze Sequence Matching (GSM), Future Action Prediction (FAP), Gaze-Triggered Alert (GTA), and Object Appearance Alert (OAA).

For each target event, only source events from the same video that occur earlier in time are eligible for retrieval. Retrieval is performed once using fixed source-event and target-query representations defined independently of the generated persistent representations. Generated Free-form or EB text is never used for retrieval, and retrieval has no access to future frames, ground-truth labels, or downstream outcomes. Source and target retrieval texts are encoded with \texttt{BAAI/bge-small-en-v1.5}~\cite{xiao2024c}, and eligible source events are ranked by cosine similarity. The top $k=3$ events are presented in chronological order and reused unchanged across all representation conditions.

\subsection{Representation Conditions}

We compare three representations of the same retrieved source events. \textbf{Answer Only} retains the model's direct source answer. \textbf{Free-form Interpretation} expresses the source information as a fluent higher-level interpretation without explicitly separating observation, inference, or uncertainty. \textbf{Evidence-Bounded Interpretation (EB)} separates retained content into \emph{Observed}, for content the model treats as directly supported by the source-time evidence; \emph{Inferred}, for higher-level interpretation; \emph{Uncertain}, for unresolved or weakly supported claims; and \emph{Temporal Scope}, for when the statement applies.

Free-form Interpretation and EB are generated by the same Qwen3.5-9B model~\cite{qwen3.5} from identical source-time inputs: gaze/video evidence, the source question, and the model's frozen direct source answer. Direct source answers are generated once before the representation comparison and are not regenerated across conditions. Source ground truth is not used for representation generation. No representation has access to future frames, target-time evidence, target labels, or downstream outcomes. The primary EB prompt was fixed before its reported evaluation outcomes were observed. At target time, all representation conditions receive the same target input and the same retrieved source events, so the persistent representation is the manipulated variable. We additionally evaluate a \textbf{No History} condition using the same target case without retrieved historical context.

\subsection{Evaluation}

Our primary analysis uses two homogeneous diagnostic strata to isolate source-history correctness. A \emph{wrong-source} (WS) case requires all three retrieved direct source predictions to be incorrect, whereas a \emph{correct-source} (CS) case requires all three to be correct. The frozen paired analysis contains 648 WS cases from 122 unique videos and 714 CS cases from 158 unique videos; mixed histories are excluded from this diagnostic analysis but remain in ordinary full-task accuracy. Under the same fixed retrieval pipeline, 10.4\% of eligible top-3 histories are all-wrong, 61.0\% mixed, and 28.6\% all-correct. These values calibrate the evaluated benchmark setting rather than deployment prevalence.

Let $y_j^{0}$ denote the No History prediction for target case $j$, $y_j^{r}$ the prediction under historical representation $r$, and $y_j^{*}$ the target ground truth. Let $\mathcal{W}$ and $\mathcal{C}$ denote the frozen WS and CS populations. We define the wrong-source harmful-flip rate as
\[
\mathrm{Harm}(r)
=
\frac{1}{|\mathcal{W}|}
\sum_{j\in\mathcal{W}}
\mathbb{1}
\left[
y_j^{0}=y_j^{*}
\land
y_j^{r}\neq y_j^{*}
\right],
\]
which measures the fraction of WS cases in which persistent history changes a correct No History prediction into an incorrect prediction. Similarly, we define the correct-source helpful-flip rate as
\[
\mathrm{Help}(r)
=
\frac{1}{|\mathcal{C}|}
\sum_{j\in\mathcal{C}}
\mathbb{1}
\left[
y_j^{0}\neq y_j^{*}
\land
y_j^{r}=y_j^{*}
\right],
\]
which measures the fraction of CS cases in which persistent history changes an incorrect No History prediction into a correct prediction. In both analyses, the denominator remains the full frozen WS or CS population rather than only the subset eligible to flip. All representation conditions are therefore evaluated on identical paired populations. Here, ``harmful'' and ``helpful'' denote model-output correctness transitions under this benchmark definition; they are not measures of user harm or benefit. We report these influence measures separately from ordinary full-task accuracy.

For single-group proportions, we report Wilson 95\% confidence intervals where appropriate. Planned representation contrasts and the WS evidence-support contrast use video-level clustered bootstrap 95\% confidence intervals, with videos resampled as clusters and paired outcomes preserved within each resample. The inferential-scope analysis is exploratory: subgroup rates use Wilson intervals, and the Prospective $-$ Observable endpoint difference is reported descriptively rather than as a formal significance test. Percentage-point differences are computed from unrounded estimates, so subtraction of printed rates may differ by 0.01 pp. Raw paired discordance counts are descriptive only.

All primary experiments use the same \texttt{Qwen/Qwen3.5-9B} checkpoint~\cite{qwen3.5} for source prediction, representation generation, and target inference. Generation uses temperature $0.7$ with the same decoding configuration and a fixed random-seed protocol. Direct source predictions are generated once and frozen before WS/CS membership is defined; Free-form and EB representations are likewise generated once and frozen before downstream evaluation. All representation conditions are evaluated on the same paired target cases. Experiments are run on a single NVIDIA A100 80GB GPU.

\section{Results}

\subsection{Free-form Interpretation Amplifies Downstream Influence}

\begin{figure*}[t]
    \centering
    \includegraphics[width=\textwidth]{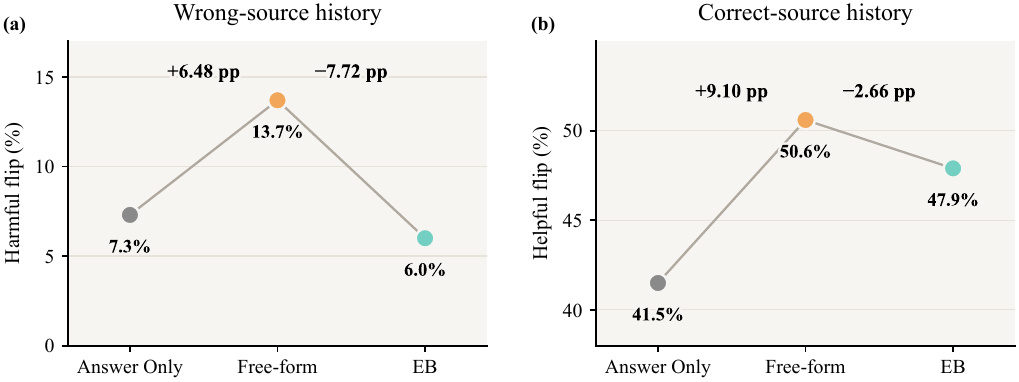}
    \caption{Changing how the same retrieved history is represented changes its downstream influence. \textbf{(a)} In the frozen wrong-source population ($n=648$), Free-form Interpretation increases harmful flips relative to Answer Only, while EB removes much of this additional harmful influence. \textbf{(b)} In the frozen correct-source population ($n=714$), Free-form Interpretation increases helpful flips, while EB retains most of this increase. Exact paired percentage-point differences and video-level clustered bootstrap 95\% confidence intervals are reported in the text.}
    \Description{Two panels compare downstream flip rates across Answer Only, Free-form Interpretation, and Evidence-Bounded Interpretation. In wrong-source histories, harmful flips rise from Answer Only to Free-form and then fall under EB. In correct-source histories, helpful flips rise from Answer Only to Free-form and remain close under EB.}
    \label{fig:representation-influence}
\end{figure*}

We first examine whether changing the persistent representation constructed from the same retrieved source events changes their downstream influence. Figure~\ref{fig:representation-influence} summarizes harmful flips in wrong-source (WS) histories and helpful flips in correct-source (CS) histories. Retrieval, source-event order, and target input are fixed across conditions.

Free-form Interpretation increases downstream influence in both directions. In the WS population, harmful flips increase from 7.3\% with Answer Only to 13.7\% with Free-form Interpretation, a paired increase of 6.48 percentage points (95\% CI [+4.17, +8.80]). In the CS population, helpful flips increase from 41.5\% to 50.6\%, a paired increase of 9.10 percentage points (95\% CI [+6.72, +11.62]). Free-form Interpretation therefore does not behave as a uniformly beneficial or harmful form of persistent context. Instead, it increases how often historical information changes later decisions in both useful and harmful directions.

Evidence-Bounded Interpretation changes this balance between harmful and helpful influence. Relative to Free-form Interpretation, EB reduces WS harmful flips from 13.7\% to 6.0\%, a decrease of 7.72 percentage points (95\% CI [-10.03, -5.56]). CS helpful flips decrease more modestly, from 50.6\% to 47.9\%, a decrease of 2.66 percentage points (95\% CI [-4.06, -1.26]). The central result is that persistent interpretation creates a two-sided trade-off: its representation changes both helpful and harmful downstream influence.

\begin{figure*}[t]
    \centering
    \includegraphics[width=\textwidth]{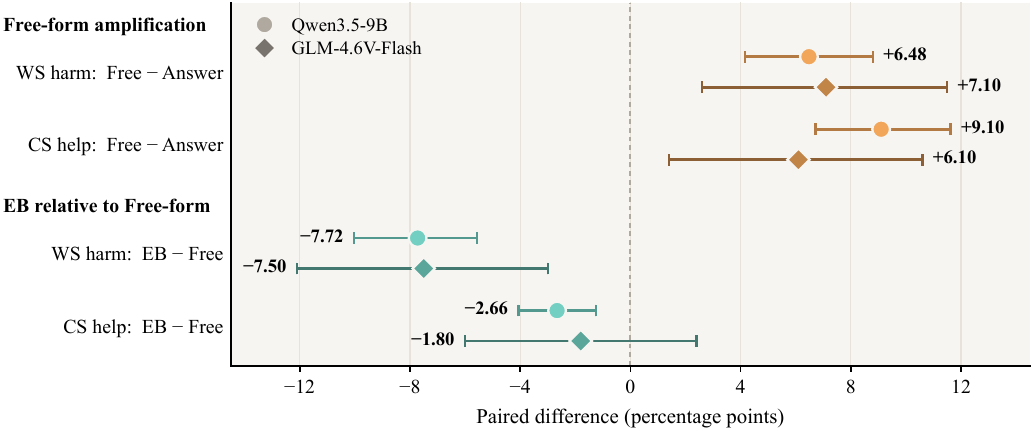}
    \caption{Cross-model robustness of the paired representation effects. Points show paired percentage-point differences and whiskers show video-level clustered bootstrap 95\% confidence intervals. Free-form Interpretation increases wrong-source harmful flips and correct-source helpful flips relative to Answer Only for both evaluated models. EB reduces wrong-source harmful influence relative to Free-form for both models, while the GLM correct-source EB--Free contrast is smaller and uncertain.}
    \Description{Horizontal forest plot comparing four paired representation effects for Qwen3.5-9B and GLM-4.6V-Flash. Both models show positive Free-form-minus-Answer effects for wrong-source harm and correct-source help, and negative EB-minus-Free effects for wrong-source harm. The GLM correct-source EB-minus-Free confidence interval crosses zero.}
    \label{fig:cross-model-robustness}
\end{figure*}

\paragraph{Cross-model robustness.}
Figure~\ref{fig:cross-model-robustness} compares the four paired representation effects across Qwen3.5-9B and GLM-4.6V-Flash~\cite{vteam2025glm45vglm41vthinkingversatilemultimodal}. GLM uses its own frozen source predictions, persistent representations, and target inference. The core amplification pattern replicates: Free-form Interpretation increases WS harm by 7.1 percentage points (95\% CI [+2.6, +11.5]) and CS help by 6.1 points (95\% CI [+1.4, +10.6]) relative to Answer Only. EB also reduces WS harm relative to Free-form by 7.5 points (95\% CI [-12.1, -3.0]). The corresponding CS difference is smaller and uncertain at -1.8 points (95\% CI [-6.0, +2.4]). The shared direction across the two evaluated models provides cross-model robustness for the core effect structure.

\subsection{When Are Persistent Interpretations Associated with More Harm?}

\begin{figure*}[t]
    \centering
    \includegraphics[width=\textwidth]{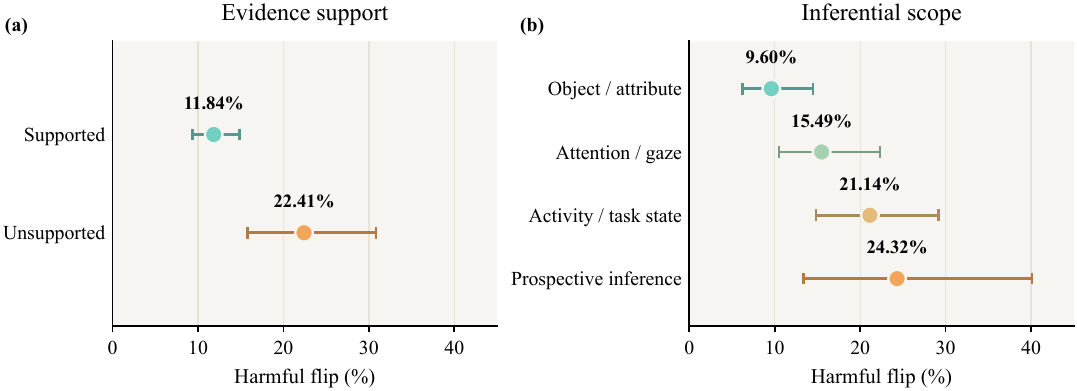}
    \caption{Properties of Free-form persistent interpretations associated with harmful downstream changes. \textbf{(a)} Within the frozen wrong-source population, unsupported interpretations have a higher harmful-flip rate than supported interpretations. \textbf{(b)} In the separate 500-case audit, harmful-flip rates are lowest for observable object or attribute statements and higher for more interpretive categories. Points and whiskers show group rates and Wilson 95\% confidence intervals. The WS support contrast is reported in the text with a video-level clustered bootstrap interval; the scope endpoint contrast is descriptive.}
    \Description{Two horizontal dot-and-whisker panels show harmful-flip rates. The first compares supported and unsupported interpretations within wrong-source histories. The second shows four inference-scope categories, from relatively observable object or attribute statements to prospective human inference.}
    \label{fig:interpretation-risk}
\end{figure*}

We examine two properties of Free-form interpretations associated with harmful changes: their support from the available earlier video and gaze evidence, and the inferential scope they add beyond relatively observable object and attention information. Figure~\ref{fig:interpretation-risk} summarizes both analyses.

\paragraph{Human annotation.}
We use two distinct annotation sets for complementary analyses. First, a detailed paired audit covers 500 unique cases from 146 unique videos, sampled without replacement from a frozen eligible pool using a fixed random seed and proportional stratification by target task. Sampling was defined independently of downstream harmful or helpful outcomes and did not condition on representation-quality or inferential-scope labels. For the inference-level analysis, annotators assign each Free-form persistent interpretation to one of four mutually exclusive levels, and we relate these labels to harmful flips under the Free-form condition. The same 500 cases are also used for a paired representation-quality comparison between Free-form Interpretation and EB, where unsupported content, contradictory content, and overcertainty are annotated as case-level binary properties for each representation. Two author annotators independently labeled all 500 cases in randomized order before discussion and without access to downstream harmful or helpful outcomes. Raw agreement was 85.0\%, with Cohen's $\kappa=0.70$ computed from the pre-adjudication labels. Disagreements were subsequently resolved through discussion, and the analyses below use adjudicated case-level labels. Because EB has a recognizable output structure, complete blinding to representation condition cannot be assumed, so this audit is used as characterization evidence rather than independent validation of the design.

Second, we use a separate binary evidence-support annotation covering the full frozen WS population ($n=648$). Here, the annotation target is the Free-form persistent interpretation. Two author annotators independently judge whether the substantive content of each interpretation is supported by the source-time video and gaze evidence in the context of the source question. When needed, the frozen direct source answer is also shown to clarify the generated statement. Annotators do not have access to source ground truth, the EB representation, target-time information, target predictions or labels, downstream harmful or helpful outcomes, or an explicit WS label. Cases are presented in randomized order. Pre-adjudication raw agreement is 91.36\%, with Cohen's $\kappa=0.71$. Disagreements are subsequently adjudicated, and the downstream analysis uses the adjudicated Supported/Unsupported labels. Full annotation protocols are provided in Appendix~\ref{app:annotation}.

\paragraph{Evidence support within wrong-source histories.}
Source correctness and representation quality are related but distinct. Across the full library of 4,513 stored interpretations, 38.3\% are anchored to an incorrect direct source answer, whereas 8.3\% are flagged by the generation pipeline's rule checks. We therefore examine evidence support within the fixed WS population rather than comparing support labels across a mixture of correct and incorrect source histories.


Within the same wrong-source stratum, Free-form harmful flips occur in 22.41\% of cases whose persistent interpretation is labeled Unsupported and 11.84\% of cases labeled Supported. The difference is +10.57 percentage points (video-level clustered bootstrap 95\% CI [+2.7, +19.3]), corresponding descriptively to a risk ratio of 1.89. The Wilson 95\% confidence intervals for the two group rates are [15.78, 30.82] and [9.37, 14.87], respectively. This association is measured within a stratum in which source correctness is fixed to the wrong-source condition.

\paragraph{Inferential scope.}
Evidence support does not capture how far a stored statement goes beyond relatively observable information. For this analysis, we use a study-specific operational taxonomy with four mutually exclusive categories: \emph{Observable object / attribute}, \emph{Attention / gaze target}, \emph{Activity / task state}, and \emph{Prospective human inference}. We compare harmful-flip rates across these categories under the Free-form condition.

Free-form harmful-flip point estimates increase from 9.60\% for observable object or attribute statements to 15.49\% for attention or gaze-target statements, 21.14\% for activity or task-state interpretations, and 24.32\% for prospective human inference. The descriptive endpoint difference between prospective human inference and observable object or attribute statements is 14.73 percentage points. Their Wilson 95\% confidence intervals are [13.36, 40.12] and [6.23, 14.50], respectively. We use this pattern as an associative characterization of inferential scope. The prospective category is small ($n=37$) and has a correspondingly wide confidence interval. The pattern is consistent with a distinction between claims tied closely to visible objects or attention and claims that extend to activity or possible future states.

These findings associate harmful downstream changes not only with source correctness, but also with evidence support and the inferential scope of the stored interpretation. This motivates testing a representation that keeps observation, inference, uncertainty, and temporal scope separate when history is reused.

\subsection{Evidence-Bounded Interpretation Reduces Harmful Changes}

We evaluate Evidence-Bounded Interpretation (EB) as a concrete response to the preceding diagnosis. EB preserves epistemic distinctions in persistent context rather than blending supported, inferred, uncertain, and temporally scoped content into one fluent statement. The primary EB prompt was fixed before its reported evaluation outcomes were observed.

\begin{figure*}[t]
    \centering
    \includegraphics[width=0.6\textwidth]{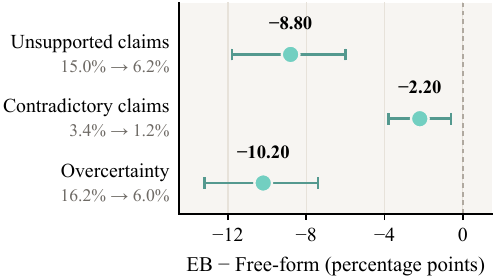}
    \caption{Changes in annotated persistent-content properties under EB relative to Free-form Interpretation in the paired 500-case audit. Points show paired percentage-point differences and whiskers show video-level clustered bootstrap 95\% confidence intervals. Gray labels show the corresponding Free-form-to-EB rates. These measures characterize specific properties of the persistent representations rather than overall representation quality.}
    \Description{Horizontal forest plot showing EB-minus-Free-form differences for unsupported claims, contradictory claims, and overcertainty. All three estimates are negative and their confidence intervals remain below zero. Absolute Free-form-to-EB rates are displayed beside each property.}
    \label{fig:eb-representation-quality}
\end{figure*}

\paragraph{EB reduces unsupported and overcertain content.}
Figure~\ref{fig:eb-representation-quality} reports the paired 500-case audit. Relative to Free-form Interpretation, EB reduces unsupported claims by 8.80 percentage points (95\% CI [-11.80, -6.00]), contradictory claims by 2.20 points (95\% CI [-3.80, -0.60]), and overcertainty by 10.20 points (95\% CI [-13.20, -7.40]). The corresponding absolute rates are shown in the figure.

EB therefore changes properties of the stored content, not only its organization. The largest reductions are in unsupported claims and overcertainty. EB is evaluated here as a complete representation design.

\paragraph{EB shifts the balance between harmful and helpful influence.}
As shown in Figure~\ref{fig:representation-influence}, EB substantially reduces WS harmful influence relative to Free-form Interpretation, while the decrease in CS helpful influence is much smaller. We therefore characterize EB as changing the balance between harmful and helpful influence rather than dominating Free-form Interpretation on every outcome.

\paragraph{The effect is consistent across target tasks.}
All four primary paired comparisons have the same direction on each of the six target tasks. Detailed task-level results are reported in Appendix~\ref{app:taskwise} as breadth evidence for the aggregate paired effects.

\subsection{Intervening on Persisted Content}

\paragraph{Correcting persisted content recovers harmful cases.}
We intervene on 117 cases from 58 unique videos that produced a harmful downstream flip under the frozen original Free-form condition, defined before any counterfactual outcome was observed: the No History prediction is correct, while the prediction with the original Free-form history is incorrect. For each case, we select the source entry with the highest similarity score under the fixed representation-independent retrieval ranking as the intervention target, before the retrieved top-$3$ entries are reordered chronologically for target inference. This selected entry is fixed across all interventions. Neutral replaces only this entry with semantically neutral wording of comparable length, Delete removes only this entry, and Oracle Historical Correction corrects only this entry using source-time evidence and, when needed, source ground truth. The other two retrieved entries and the target input remain unchanged. None of the interventions has access to future frames, target-time evidence, target labels, or downstream outcomes.

\begin{table}[t]
\centering
\caption{Recovery after intervening on 117 harmful cases. Oracle historical correction uses source-time evidence and source ground truth when needed, without access to future frames, target-time evidence, target labels, or downstream outcomes.}
\label{tab:counterfactual-recovery}
\begin{tabular}{lr}
\toprule
Historical intervention & Recovery \\
\midrule
Neutral length-matched edit & 11.11\% \\
Delete selected entry & 25.64\% \\
Oracle historical correction & \textbf{41.03\%} \\
\bottomrule
\end{tabular}
\end{table}

Oracle historical correction recovers 41.03\% of harmful cases, compared with 25.64\% after removing the selected historical entry and 11.11\% after a neutral length-matched edit. Oracle correction exceeds deletion by 15.38 percentage points (95\% CI [+8.55, +22.22]) and the neutral edit by 29.91 percentage points (95\% CI [+21.37, +39.32]). These interventions show that the semantic content of the selected persistent entry contributes to later failures. Because oracle correction may use source ground truth, we use it as a mechanism intervention. Appendix~\ref{app:case-study} provides one qualitative example tracing this process from source evidence to later outcome.

\subsection{Overall Task Performance and Proactive Outputs}

\paragraph{The trade-off is task dependent.}
Harmful and helpful flips isolate controlled changes relative to No History; Table~\ref{tab:full-task-accuracy} reports ordinary full-task accuracy.

\begin{table}[t]
\centering
\caption{Full-task accuracy under No History, Free-form Interpretation, and Evidence-Bounded Interpretation.}
\label{tab:full-task-accuracy}
\begin{tabular}{lrrr}
\toprule
Task & No History & Free-form & Evidence-Bounded \\
\midrule
OTP & 41.9\% & \textbf{48.8\%} & 48.0\% \\
SR  & \textbf{53.6\%} & 43.1\% & 50.9\% \\
GSM & 53.8\% & \textbf{79.6\%} & 76.7\% \\
FAP & 29.8\% & \textbf{41.5\%} & 39.2\% \\
GTA & 63.5\% & 63.8\% & \textbf{65.5\%} \\
OAA & \textbf{69.4\%} & 65.8\% & 69.0\% \\
\midrule
Six-task macro & 52.0\% & 57.1\% & \textbf{58.2\%} \\
\bottomrule
\end{tabular}
\end{table}

EB retains most of the Free-form gains on OTP, GSM, and FAP and recovers much of its degradation on SR and OAA. Six-task macro accuracy is 58.2\% with EB, 57.1\% with Free-form, and 52.0\% with No History. EB nevertheless remains below No History on SR (50.9\% vs.\ 53.6\%) and slightly below on OAA (69.0\% vs.\ 69.4\%), showing that EB does not make persistent history useful for every target behavior.

\paragraph{Persistent representation also changes system-triggered outputs.}
GTA and OAA extend the analysis from passive answers to outputs in which the assistant must decide when to trigger an alert based on gaze or object appearance. EB reduces harmful changes on both tasks: by 7.74 percentage points on GTA (95\% CI [-10.62, -4.87]) and by 8.64 points on OAA (95\% CI [-11.68, -5.61]). The corresponding gain-rate differences are smaller: -1.55 points on GTA (95\% CI [-3.54, +0.22]) and +0.47 points on OAA (95\% CI [-2.10, +3.04]). Thus, the influence of persistent representation reaches not only later answers, but also benchmark decisions about when the assistant acts. Appendix Figure~\ref{fig:proactive-effects} visualizes these paired effects.

\section{Discussion}

\subsection{Persistence Changes What Interpretation Does}

Persistence changes the role of a machine interpretation. At source time, an interpretation is a statement about an earlier event. Once retained, it becomes part of the system state used to reason about what happens next. Our two-sided amplification result shows why this distinction matters: Free-form Interpretation makes correct history more capable of changing later decisions in a helpful direction, but it also makes wrong history more capable of changing them in a harmful direction. Interpretation is therefore not only a way of describing prior evidence. When it persists, it becomes a mechanism through which the system assigns influence to its own earlier understanding.

This matters for interactive assistants because later behavior is not limited to passive question answering. The evaluated target tasks include future-action prediction and system-triggered alerts based on gaze or object appearance. Our proactive results show that the representation of earlier interpretation changes these later system-triggered outputs as well. A statement that begins as an interpretation of one moment can therefore become part of the context behind what the assistant predicts, recalls, or decides to act on later.

Gaze makes this transition especially visible because it creates an epistemic boundary between relatively direct behavioral evidence and broader interpretation. Visible objects and attended regions provide evidence about what a person saw or looked at, while activity and possible future behavior require additional inference. Within wrong-source histories, unsupported interpretations are associated with substantially more harmful changes, and the separate scope analysis shows higher harmful-flip rates for broader interpretive categories. When such distinctions are flattened into one fluent persistent statement, later reasoning receives the interpretation without an equally explicit account of how far it extends beyond the underlying evidence.



\subsection{Design Principles for History-Aware Assistants}

Our findings point to three design choices that become important when AI systems retain interpretations about people across an unfolding interaction. Prior work on uncertainty and human-AI decision making has largely considered how the status of model output is communicated to people~\cite{salimzadeh2024dealing,schafer2025ai,kim2024m,ma2024you}. Our results show a complementary issue: epistemic status also matters in the information an AI system carries forward for its own later reasoning.

\paragraph{Preserve evidence status when interpretation persists.}
A persistent representation should retain the distinction between what the system observed, what it inferred, and what remains uncertain~\cite{jin2026mitigating,xu2026memory,liao2026belief,zhang2026agentic}. For example, retaining that a person is ``preparing to cut'' is different from retaining that the person looked at a knife and tomato and that preparing to cut is an inference from those observations. Free-form Interpretation tends to collapse these roles into one fluent statement. EB provides one tested way to keep them available to later reasoning.

\paragraph{Evaluate persistent representations at the point of reuse.}
The quality of persistent interpretation cannot be judged only by whether a stored statement appears plausible in isolation. Its consequence is realized when that statement is retrieved and changes a later decision. Our counterfactual intervention shows that correcting the semantic content of a harmful historical entry recovers more cases than deleting it or replacing it with neutral text. Persistent representations should therefore be evaluated by how their later reuse changes system behavior, including predictions and system-triggered outputs.

\paragraph{Separate persistence from reuse.}
Retaining an interpretation and using it for a later decision are distinct design choices. Our full-task results show that history is useful for some target tasks but not for every target behavior. A system may therefore have reason to preserve an earlier interpretation without making it part of every subsequent decision. This separates three questions that are often collapsed in history-aware systems: what should persist, how it should be represented, and when it should be reused~\cite{xiong2026memory,pan2025secom,yan2026memory}.

\subsection{Limitations}

Our experiments evaluate persistent interpretation in a controlled gaze-aware benchmark. The measured outcomes are changes in model correctness, while user-facing outcomes such as trust, comfort, and perceived appropriateness remain outside the present evaluation.

The primary analysis uses Qwen3.5-9B on STREAMGAZE, with GLM-4.6V-Flash providing a second-model replication of the core effect structure. WS and CS are homogeneous diagnostic strata: under the evaluated retrieval pipeline, all-wrong, mixed, and all-correct histories account for 10.4\%, 61.0\%, and 28.6\% of eligible histories, respectively. Mixed histories remain included in ordinary full-task accuracy.

The human analyses characterize associations between persistent content and downstream behavior. The evidence-support analysis fixes source correctness within the wrong-source population, while the inference-scope analysis includes a small prospective-inference group and correspondingly wide uncertainty. Both annotation studies use two author annotators who label cases independently before adjudication and without access to downstream outcomes; representation condition remains visually recognizable in the paired EB audit.

Finally, the main manipulation evaluates persistent representation as a whole, including both semantic elaboration and textual organization. EB is likewise evaluated as a complete representation design. The neutral intervention isolates a role for semantic content, while oracle historical correction provides intervention evidence using source-time ground truth.

\section{Conclusion}

We show that turning behavioral evidence into persistent machine interpretation is not a neutral way of carrying history forward. Free-form Interpretation amplifies the influence of earlier information in both directions: correct history becomes more capable of producing helpful downstream changes, while wrong history becomes more capable of producing harmful ones. Evidential support and inferential scope further characterize which persisted interpretations are associated with greater harmful influence, and the same core amplification structure appears across both evaluated models.

These findings shift the design question from whether an assistant should interpret behavior to how its interpretations should persist. Evidence-Bounded Interpretation shows that keeping observation, inference, uncertainty, and temporal scope distinguishable can substantially reduce harmful influence from wrong histories while retaining most of the useful influence from correct ones. More broadly, when AI systems carry interpretations of people across an interaction, they also carry forward assumptions about what was observed, what was inferred, and what was uncertain. Those distinctions should remain part of the persistent context used to shape later assistance.

\section*{Ethical Statement}
This study analyzes how model-generated interpretations of behavioral evidence influence later model decisions in a controlled gaze-aware benchmark. It does not measure whether a model understands a person's intentions, activities, or future behavior, and the reported harmful and helpful changes refer only to model-output correctness rather than user harm or benefit. The findings should therefore not be used to justify profiling or making consequential judgments about real individuals based on gaze or model-generated interpretations. All experiments use the existing STREAMGAZE benchmark, and the human annotation reported in this paper was conducted by the authors on benchmark evidence and model outputs.

\bibliographystyle{ACM-Reference-Format}
\bibliography{sample-base}

\clearpage

\appendix
\section{Experimental Protocol and Reproducibility}
\label{app:experimental}

\paragraph{Source and target events.}
We use four STREAMGAZE tasks as sources of earlier gaze-conditioned events: Object Identification Easy (OI-E), Object Identification Hard (OI-H), Object Attribute Recognition (OAR), and Non-Fixated Object Identification (NFI). The resulting source library contains 4,513 entries anchored to direct source predictions. We evaluate downstream influence on six target tasks: Object Transition Prediction (OTP), Scene Recall (SR), Gaze Sequence Matching (GSM), Future Action Prediction (FAP), Gaze-Triggered Alert (GTA), and Object Appearance Alert (OAA). Source and target events come from the same video, and a source event is eligible only when it occurs before the target event.

\paragraph{Fixed retrieval representation.}
Retrieval uses fixed textual event representations encoded with \texttt{BAAI/bge-small-en-v1.5}~\cite{xiao2024c} and ranked by cosine similarity. Each source event is represented by concatenating its source question with the model's frozen direct source answer, while the target retrieval query contains only the target question. Source and target texts are encoded in the same embedding space. Generated Free-form or EB text is never used in either retrieval representation, so retrieval is fixed independently of the persistent representation being evaluated.

\paragraph{Time-causal retrieval protocol.}
For each target event, retrieval is performed once over eligible earlier events from the same video. The retrieval stage has no access to future frames, source or target ground truth, target labels, downstream predictions, or downstream outcomes. We select the top $k=3$ source events according to cosine similarity and reorder the selected events chronologically before target inference. The same three events and the same chronological order are reused unchanged across Answer Only, Free-form Interpretation, and Evidence-Bounded Interpretation.

\paragraph{Primary-model generation and freezing protocol.}
For the primary Qwen3.5-9B experiment, all representation conditions use the same checkpoint, decoding configuration, and fixed random-seed protocol. Direct source predictions are generated once and then frozen. Wrong-source and correct-source membership is defined once from these frozen direct source predictions and remains unchanged throughout the representation comparison. Free-form and EB representations are also generated once from the frozen source-time inputs and are frozen before downstream evaluation. No representation condition regenerates source answers or changes WS/CS membership, and all representation conditions are evaluated on the same paired target cases.

\paragraph{Representation generation.}
Free-form Interpretation and EB receive identical source-time inputs: the causal gaze/video evidence available at the source event, the source question, and the model's frozen direct source answer. Source ground truth is not used for the primary representation generation. Representation generation has no access to future frames, target-time evidence, target labels, or downstream outcomes. The primary EB prompt was fixed before its reported evaluation outcomes were observed.

\paragraph{Cross-model replication protocol.}
We repeat the complete representation pipeline with \texttt{zai-org/GLM-4.6V-Flash}~\cite{vteam2025glm45vglm41vthinkingversatilemultimodal}. GLM independently generates its own direct source predictions, which are generated once and frozen before defining GLM-specific wrong-source and correct-source populations. Free-form and EB representations are then generated by GLM from its own frozen source-time inputs and are frozen before downstream evaluation. Retrieval follows the same causal and representation-independent protocol as in the primary experiment: each source retrieval text concatenates the source question with GLM's frozen direct source answer, the target retrieval text contains the target question, and both are encoded with \texttt{BAAI/bge-small-en-v1.5}~\cite{xiao2024c}. For each target, the top-$k=3$ eligible earlier events are retrieved once and reused in the same chronological order across Answer Only, Free-form Interpretation, and EB. GLM then performs target inference for all representation conditions. This replication tests whether the primary amplification and mitigation patterns remain visible when source prediction, representation generation, and target inference are all performed by a second model family.

\paragraph{Target inference.}
At target time, all history-aware representation conditions receive the same causal target input and the same three retrieved earlier source events. Only the textual representation of those earlier events changes across Answer Only, Free-form Interpretation, and EB. No History uses the same target case without retrieved historical context. The target model has no access to frames occurring after the target time.

\paragraph{Primary-model implementation.}
The same \texttt{Qwen/Qwen3.5-9B} checkpoint~\cite{qwen3.5} is used for source prediction, representation generation, and target inference in the primary experiment. Generation uses temperature $0.7$ with the same decoding configuration and fixed random-seed protocol across conditions. All primary experiments are run on a single NVIDIA A100 80GB GPU.

\section{Representation Generation Specification}
\label{app:prompts}

Free-form Interpretation and Evidence-Bounded Interpretation are generated from the same source-time inputs. In both conditions, the model receives the causal source-time gaze/video evidence, the source question, and the frozen direct source answer. Representation generation does not use source ground truth, future frames, target-time evidence, target labels, downstream predictions, or downstream outcomes. The generation rules are fixed across all evaluated cases. The following specifications summarize the information and output constraints that distinguish the two representation conditions.

\subsection{Free-form Interpretation}

The Free-form condition asks the model to convert the source-time information into a concise natural-language interpretation without explicitly separating observation, inference, uncertainty, or temporal scope. Its generation specification is:

\begin{verbatim}
You are given an earlier event from an egocentric
gaze-aware video stream.

Use only the information available at this earlier event:
- the source-time video and gaze evidence,
- the source question, and
- the model's source answer.

Write one concise natural-language interpretation of what
this evidence suggests about the person's current activity,
situation, or plausible next behavior.

You may connect the observed objects, gaze information,
and surrounding context into a higher-level interpretation.
Do not refer to any later event, future video evidence,
target question, target answer, or ground-truth label.

Output only the interpretation.
\end{verbatim}

This condition intentionally permits observation and higher-level inference to appear in one fluent persistent statement. It does not explicitly require the generated text to identify which parts are directly observed, inferred, or uncertain.

\subsection{Evidence-Bounded Interpretation}

The EB condition receives the same source-time inputs but requires the status of retained information to remain explicit. Its generation specification is:

\begin{verbatim}
You are given an earlier event from an egocentric
gaze-aware video stream.

Use only the information available at this earlier event:
- the source-time video and gaze evidence,
- the source question, and
- the model's source answer.

Represent the event using exactly four fields:

Observed: State only what is directly supported by the
available source-time video and gaze evidence.

Inferred: State the most reasonable interpretation that can
be drawn from the observed evidence. Do not present an
inference as an observation.

Uncertain: State what remains ambiguous or is not supported
well enough to treat as established, especially claims about
the person's next action, intention, or future state.

Temporal Scope: State whether the retained information
describes the earlier observed event, an interpretation of
that event, or a possible future state.

Do not use later video evidence, a target question, a target
answer, source or target ground truth, or downstream
outcomes.

Output exactly these four fields and no additional text.
\end{verbatim}

EB does not prohibit interpretation. Its purpose is to keep directly supported evidence, model inference, uncertainty, and temporal scope distinguishable when the generated content becomes persistent context. The primary EB prompt was fixed before its reported downstream evaluation outcomes were observed.

\paragraph{Generation-rule violations.}
The reported 8.3\% rate refers to 373 of the 4,513 Qwen library entries flagged by the task-specific rule checks recorded in the generation pipeline. The pipeline uses JSON parsing, rule-based post-checks, and one corrective retry after a detected violation. This diagnostic flag is separate from direct source-answer correctness.

\section{Human Annotation Protocols}
\label{app:annotation}

\subsection{Detailed Paired Audit}

We conducted a detailed audit of 500 unique cases from 146 unique videos, sampled without replacement from a frozen eligible interpretation pool using a fixed random seed and proportional stratification by target task. The sampling procedure was defined before annotation and did not condition on downstream harmful or helpful outcomes, representation-quality labels, or inferential-scope categories. A case consists of the source-time evidence and the corresponding persistent representation to be evaluated. Two author annotators independently labeled all 500 cases in randomized order before any discussion, yielding 1,000 pre-adjudication judgments. The annotation interface did not expose downstream harmful or helpful outcomes. For this detailed 500-case audit, the reported pre-adjudication agreement statistics are 85.0\% raw agreement and Cohen's $\kappa=0.70$. Disagreements were subsequently resolved through discussion, and all reported prevalence and characterization analyses use the adjudicated case-level labels.

The detailed audit supports two separate analyses. First, the Free-form persistent interpretation is assigned to one of four mutually exclusive inferential-scope categories. These labels are later related to harmful flips under the Free-form condition. Second, the same 500 source cases support a paired representation-quality comparison between Free-form Interpretation and EB, where unsupported content, contradictory content, and overcertainty are coded as binary properties for each representation.

Because EB has a recognizable four-field structure, complete blinding to representation condition cannot be guaranteed in the paired representation-quality audit. We therefore use these annotations as characterization evidence rather than as independent validation of the EB design. This limitation does not affect the primary downstream behavioral comparisons, which are computed directly from model outputs.

\subsection{Representation-Quality Labels}

For each representation, annotators evaluate three binary properties.

\emph{Unsupported content} is present when the persistent representation contains a substantive claim that is not supported by the available source-time video and gaze evidence.

\emph{Contradictory content} is present when the persistent representation contains a substantive claim that conflicts with the available source-time evidence.

\emph{Overcertainty} is present when uncertain or inferential content is expressed with stronger certainty than the source-time evidence supports.

These properties characterize the generated persistent representation and are distinct from whether the frozen direct source answer itself is correct.

\subsection{Inferential-Scope Labels}

Each Free-form persistent interpretation is assigned to exactly one of four categories according to the broadest type of human-related claim expressed in the persistent text.

\emph{Observable object / attribute} covers statements restricted to visible objects or their observable properties.

\emph{Attention / gaze target} covers statements about what object or region the person is attending to.

\emph{Activity / task state} covers interpretations of what activity the person is performing or the state of the ongoing task.

\emph{Prospective human inference} covers claims about a possible future action, intention, or future state of the person.

The final adjudicated distribution is 198 Observable object / attribute cases, 142 Attention / gaze target cases, 123 Activity / task state cases, and 37 Prospective human inference cases. These labels are used for associative characterization only. They are not treated as interventions on inferential scope.

\subsection{Wrong-Source Evidence-Support Annotation}

We use a separate binary evidence-support annotation covering the full frozen wrong-source population ($n=648$). This annotation set is distinct from the detailed 500-case audit. The annotation target is the Free-form persistent interpretation.

For each case, two author annotators independently judge whether the substantive content of the Free-form interpretation is supported by the source-time video and gaze evidence in the context of the source question. When needed, the frozen direct source answer is also shown to clarify the referent or meaning of the generated statement. \emph{Supported} indicates that the substantive content of the persistent interpretation can be justified from the available source-time materials. \emph{Unsupported} indicates that the interpretation adds substantive content that cannot be justified from those materials.

Cases are presented in randomized order. Annotators do not have access to source ground truth, the EB representation, target-time information, target predictions or labels, downstream harmful or helpful outcomes, or an explicit indication that a case belongs to the WS population. Before adjudication, the two annotators agree on 592 of 648 cases (91.36\%), with Cohen's $\kappa=0.71$. The remaining 56 disagreements are subsequently adjudicated, yielding 532 Supported and 116 Unsupported cases. Among these adjudicated groups, 63 Supported cases and 26 Unsupported cases produce harmful flips under Free-form Interpretation, corresponding to harmful-flip rates of 11.84\% and 22.41\%, respectively.

\section{Metrics and Statistical Analysis}
\label{app:statistics}

\subsection{Analysis Populations}

The analyses use several populations for different questions. These populations should not be summed: some are overlapping subsets, while others correspond to separate annotation or task-specific analyses. Table~\ref{tab:analysis-populations} summarizes their roles.

\begin{table}[htbp]
\centering
\caption{Analysis populations used throughout the study.}
\label{tab:analysis-populations}
\begin{tabular}{lrl}
\toprule
Population & $n$ & Primary use \\
\midrule
Stored source interpretations & 4,513 & Library-level context \\
Wrong-source histories & 648 & Harmful-flip analysis \\
Correct-source histories & 714 & Helpful-flip analysis \\
Detailed paired audit & 500 & Scope and representation quality \\
WS support annotation & 648 & Evidence support within WS \\
Harmful-case interventions & 117 & Counterfactual mechanism \\
GTA proactive cases & 452 & Proactive-output analysis \\
OAA proactive cases & 428 & Proactive-output analysis \\
\bottomrule
\end{tabular}
\end{table}

\subsection{Frozen Wrong-Source and Correct-Source Populations}

A wrong-source (WS) history requires all three retrieved direct source predictions to be incorrect. A correct-source (CS) history requires all three retrieved direct source predictions to be correct. The final paired analysis contains $|\mathcal{W}|=648$ WS cases and $|\mathcal{C}|=714$ CS cases. Histories containing both correct and incorrect retrieved source predictions are excluded from the primary WS/CS analysis but remain included in ordinary full-task accuracy.

Let $y_j^{0}$ denote the No History prediction for target case $j$, $y_j^{r}$ the prediction under persistent representation $r$, and $y_j^{*}$ the target ground truth. The harmful-flip rate is
\[
\mathrm{Harm}(r)
=
\frac{1}{|\mathcal{W}|}
\sum_{j\in\mathcal{W}}
\mathbb{1}
\left[
y_j^{0}=y_j^{*}
\land
y_j^{r}\neq y_j^{*}
\right].
\]
The helpful-flip rate is
\[
\mathrm{Help}(r)
=
\frac{1}{|\mathcal{C}|}
\sum_{j\in\mathcal{C}}
\mathbb{1}
\left[
y_j^{0}\neq y_j^{*}
\land
y_j^{r}=y_j^{*}
\right].
\]

These definitions use the full frozen WS and CS populations as their respective denominators. No History correctness therefore determines whether an item contributes a harmful or helpful flip, but it does not redefine the WS or CS denominator.

\subsection{Confidence Intervals}

For single-group proportions, we report Wilson 95\% confidence intervals where appropriate. Paired representation, counterfactual-intervention, and proactive-output differences use video-level clustered bootstrap 95\% confidence intervals. Videos are resampled as clusters so that all cases from the same video remain together, while paired condition outcomes remain paired within each resample. The primary WS and CS populations span 122 and 158 unique videos, respectively; the detailed 500-case audit spans 146 videos, and the 117-case counterfactual intervention spans 58 videos. The WS evidence-support analysis uses the same 122-video WS population and the same clustered procedure for the Unsupported $-$ Supported contrast. The inferential-scope analysis is exploratory: subgroup rates use Wilson intervals, and the Prospective $-$ Observable endpoint difference is reported descriptively. Raw paired discordance counts $b/c$ are descriptive only.

\subsection{Raw Paired Discordance}

\begin{table}[t]
\centering
\caption{Raw paired discordance counts for the primary representation comparisons. These counts are descriptive only.}
\label{tab:paired-discordance}
\begin{tabular}{lrr}
\toprule
Comparison & $b$ & $c$ \\
\midrule
WS harm, Free $-$ Answer & 50 & 8 \\
CS help, Free $-$ Answer & 74 & 9 \\
WS harm, EB $-$ Free & 5 & 55 \\
CS help, EB $-$ Free & 5 & 24 \\
\bottomrule
\end{tabular}
\end{table}

\section{Detailed Task-Level Paired Results}
\label{app:taskwise}

Table~\ref{tab:taskwise-final} reports the final task-level paired results across OTP, SR, GSM, FAP, GTA, and OAA. The same directional structure appears across all six target tasks: Free-form Interpretation increases WS harm relative to Answer Only, EB decreases WS harm relative to Free-form Interpretation, Free-form Interpretation increases CS help relative to Answer Only, and EB decreases CS help relative to Free-form Interpretation.

\begin{table*}[t]
\centering
\caption{Detailed task-level paired results under the final frozen WS/CS analysis. Differences are reported in percentage points.}
\label{tab:taskwise-final}
\begin{tabular}{lrrr rr rrr rr}
\toprule
& \multicolumn{5}{c}{Wrong-source harm} & \multicolumn{5}{c}{Correct-source help} \\
\cmidrule(lr){2-6} \cmidrule(lr){7-11}
Task & $n$ & Answer & Free & EB & $\Delta_{\mathrm{F-A}}$ & $n$ & Answer & Free & EB & $\Delta_{\mathrm{F-A}}$ \\
\midrule
OTP & 118 & 4.2\% & 14.4\% & 5.9\% & +10.17 & 124 & 45.2\% & 52.4\% & 50.0\% & +7.26 \\
SR  & 96  & 8.3\% & 11.5\% & 7.3\% & +3.12  & 109 & 45.0\% & 55.0\% & 52.3\% & +10.09 \\
GSM & 112 & 8.0\% & 15.2\% & 8.9\% & +7.14  & 131 & 36.6\% & 44.3\% & 40.5\% & +7.63 \\
FAP & 103 & 7.8\% & 15.5\% & 5.8\% & +7.77  & 115 & 37.4\% & 47.0\% & 44.3\% & +9.57 \\
GTA & 121 & 9.1\% & 10.7\% & 3.3\% & +1.65  & 118 & 44.1\% & 59.3\% & 55.9\% & +15.25 \\
OAA & 98  & 6.1\% & 15.3\% & 5.1\% & +9.18  & 117 & 41.0\% & 46.2\% & 45.3\% & +5.13 \\
\bottomrule
\end{tabular}
\end{table*}

For WS harm, the EB-minus-Free differences for OTP, SR, GSM, FAP, GTA, and OAA are $-8.47$, $-4.17$, $-6.25$, $-9.71$, $-7.44$, and $-10.20$ percentage points, respectively. For CS help, the corresponding EB-minus-Free differences are $-2.42$, $-2.75$, $-3.82$, $-2.61$, $-3.39$, and $-0.85$ percentage points. We use this task-level analysis as breadth evidence and do not require every task-level contrast to be individually significant.

\section{Counterfactual Intervention Details}
\label{app:counterfactual}

We intervene on the persisted historical interpretation for the same 117 cases from 58 unique videos that produce harmful downstream changes under the frozen original Free-form reference condition. Each case has one fixed selected historical entry as the intervention target. Neutral editing, deletion, and oracle correction all operate on this same entry, while the other two retrieved entries and the target input remain unchanged. The \emph{neutral length-matched edit} replaces the source-specific semantic content of the selected entry with semantically neutral wording of comparable length while keeping a textual entry present. The \emph{deletion} condition removes the selected entry. The \emph{oracle historical correction} replaces unsupported or incorrect content in the selected entry using source-time evidence and source ground truth when needed. The 117 intervention cases are defined before any counterfactual intervention outcome is observed, and target inference uses the same decoding configuration and fixed random-seed protocol as the original evaluation. Oracle correction has no access to future frames, target-time evidence, target labels, or downstream outcomes. A case is counted as recovered when the intervention changes the originally harmful target prediction back to the correct target answer.

\begin{table}[t]
\centering
\caption{Recovery rates under counterfactual interventions on 117 harmful cases.}
\label{tab:counterfactual-full}
\begin{tabular}{lr}
\toprule
Intervention & Recovery \\
\midrule
Neutral length-matched edit & 11.11\% \\
Delete selected entry & 25.64\% \\
Oracle historical correction & \textbf{41.03\%} \\
\bottomrule
\end{tabular}
\end{table}

\begin{table}[t]
\centering
\caption{Paired differences between counterfactual interventions.}
\label{tab:counterfactual-paired}
\begin{tabular}{lrr}
\toprule
Comparison & $\Delta$ (pp) & 95\% CI \\
\midrule
Oracle $-$ Delete & \textbf{+15.38} & [+8.55, +22.22] \\
Oracle $-$ Neutral & \textbf{+29.91} & [+21.37, +39.32] \\
Delete $-$ Neutral & \textbf{+14.53} & [+7.69, +22.22] \\
\bottomrule
\end{tabular}
\end{table}

\section{Proactive Output Analysis}
\label{app:proactive}

The proactive GTA and OAA tasks test whether persistent historical interpretation changes later system-triggered outputs. For each proactive task $\tau$, let $\mathcal{P}_{\tau}$ denote its fixed proactive analysis population, with $|\mathcal{P}_{\mathrm{GTA}}|=452$ and $|\mathcal{P}_{\mathrm{OAA}}|=428$. Let $y_j^{0}$ denote the No History prediction, $y_j^{r}$ the prediction under representation $r$, and $y_j^{*}$ the target label. We define
\[
\mathrm{Hurt}_{\tau}(r)
=
\frac{1}{|\mathcal{P}_{\tau}|}
\sum_{j\in\mathcal{P}_{\tau}}
\mathbb{1}
\left[
y_j^{0}=y_j^{*}
\land
y_j^{r}\neq y_j^{*}
\right],
\]
and
\[
\mathrm{Gain}_{\tau}(r)
=
\frac{1}{|\mathcal{P}_{\tau}|}
\sum_{j\in\mathcal{P}_{\tau}}
\mathbb{1}
\left[
y_j^{0}\neq y_j^{*}
\land
y_j^{r}=y_j^{*}
\right].
\]
Hurt therefore captures proactive cases changed from correct to incorrect by persistent history, whereas Gain captures cases changed from incorrect to correct. These rates are computed within the fixed proactive analysis populations and are not used to reconstruct the independently evaluated full-task accuracies.

\begin{figure*}[t]
    \centering
    \includegraphics[width=0.62\textwidth]{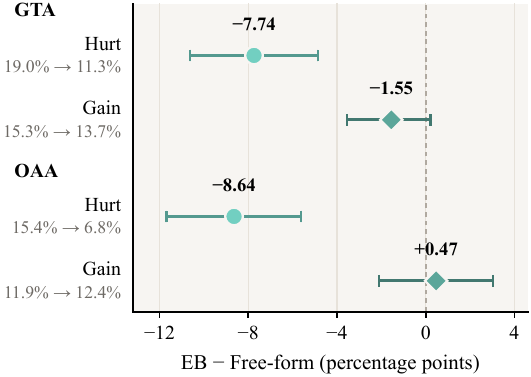}
    \caption{Paired proactive-task effects of EB relative to Free-form Interpretation. Circles denote Hurt and diamonds denote Gain. Points show percentage-point differences and whiskers show video-level clustered bootstrap 95\% confidence intervals. EB reduces Hurt on both GTA and OAA, while the corresponding Gain differences are smaller and their intervals include zero. These are benchmark model-output metrics rather than measures of user harm or benefit.}
    \Description{Horizontal forest plot showing EB-minus-Free-form effects for Hurt and Gain on GTA and OAA. The Hurt estimates for both tasks are negative with confidence intervals below zero. Both Gain confidence intervals cross zero.}
    \label{fig:proactive-effects}
\end{figure*}

Figure~\ref{fig:proactive-effects} shows substantially lower Hurt under EB on both proactive tasks, while the corresponding Gain differences are smaller and uncertain. We therefore interpret this result as a change in proactive benchmark outputs rather than evidence of user-level harm reduction.

\section{Qualitative Case Study}
\label{app:case-study}

\begin{figure*}[htbp]
    \centering
    \includegraphics[width=\textwidth]{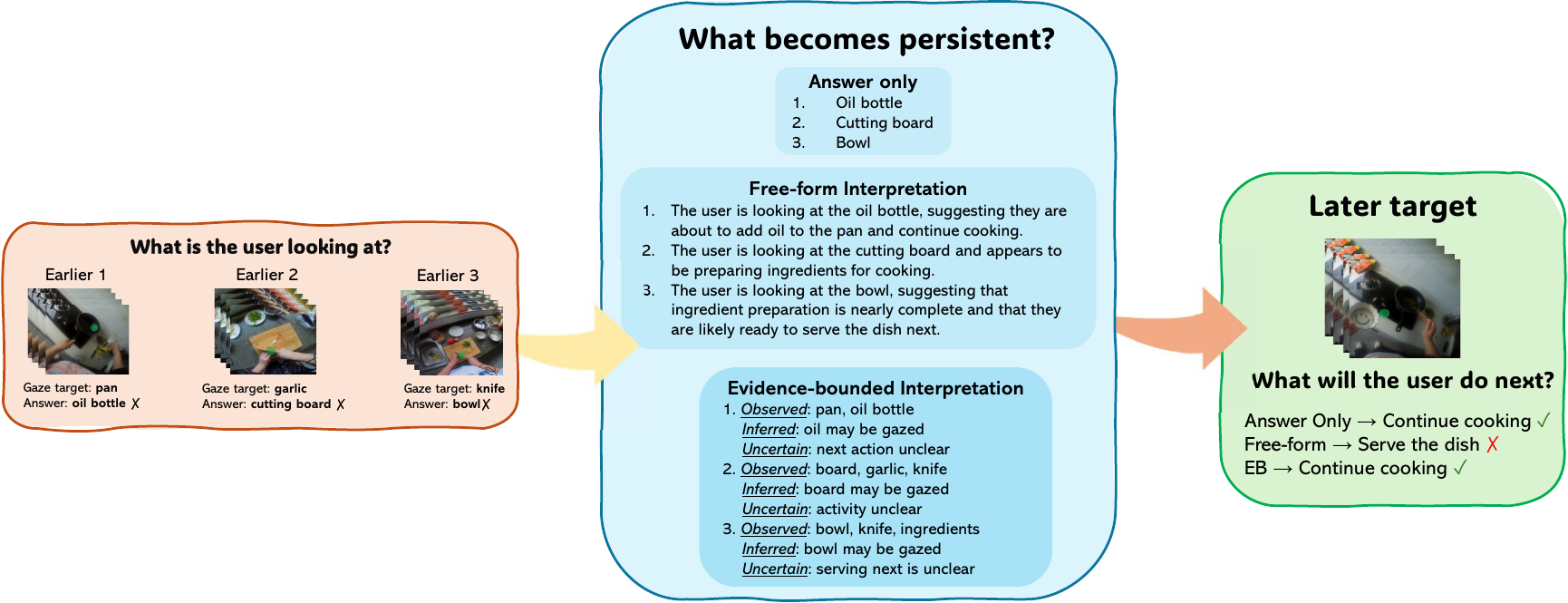}
    \caption{Illustrative case of harmful propagation and the EB response. The same three earlier source events are retained under each representation condition. The later target prediction is correct under Answer Only, changes to an incorrect prediction under Free-form Interpretation, and is correct again under EB. The Free-form history extends the earlier evidence into a fluent prospective interpretation, whereas EB keeps observation, inference, uncertainty, and temporal scope distinguishable. This example is illustrative and is not used as quantitative evidence.}
    \Description{A qualitative appendix figure showing source evidence, a Free-form persistent interpretation, an EB persistent interpretation, and their later downstream consequences.}
    \label{fig:case-study}
\end{figure*}

Figure~\ref{fig:case-study} shows one concrete chain from source evidence to persistent representation to later target behavior. The purpose of this example is illustrative: it helps explain how a fluent Free-form interpretation can propagate beyond the source-time evidence, and how EB can preserve a more bounded representation of the same earlier event.

\end{document}